\documentclass[twocolumn,floatfix,amssymb,aps,superscriptaddress,letterpager]{revtex4}
\usepackage[colorlinks=true,urlcolor=blue,citecolor=blue,linkcolor=blue]{hyperref}
\usepackage{graphicx}
\usepackage{color}
\usepackage{epsfig}
\usepackage{amsmath}
\usepackage{amssymb}
\usepackage{mathrsfs}
\usepackage{bm}
\usepackage{subfigure}
\usepackage{float}
\usepackage{url}
\usepackage{enumerate}
\usepackage{braket}
\usepackage{comment}
\usepackage{multirow}
\definecolor{violet}{rgb}{0.56,0.0,1.0}
\begin{document}

\hsize\textwidth\columnwidth\hsize\csname@twocolumnfalse\endcsname

\title{Magnetization of the spin-$1/2$ Heisenberg antiferromagnet on the triangular lattice}

\author{Qian Li}
\affiliation{School of Physical Science and Technology $\&$ Key Laboratory for Magnetism and
Magnetic Materials of the MoE, Lanzhou University, Lanzhou 730000, China}
\affiliation{Lanzhou Center for Theoretical Physics and Key Laboratory of Theoretical Physics of Gansu Province, Lanzhou University, Lanzhou 730000, China.}

\author{Hong Li}
\affiliation{Department of Physics,  Renmin University of China,  Beijing 100872,  China}

\author{Jize Zhao}
\email[]{zhaojz@lzu.edu.cn}
\affiliation{School of Physical Science and Technology $\&$ Key Laboratory for Magnetism and
Magnetic Materials of the MoE, Lanzhou University, Lanzhou 730000, China}
\affiliation{Lanzhou Center for Theoretical Physics and Key Laboratory of Theoretical Physics of Gansu Province, Lanzhou University, Lanzhou 730000, China.}

\author{Hong-Gang Luo}
\affiliation{School of Physical Science and Technology $\&$ Key Laboratory for Magnetism and
Magnetic Materials of the MoE, Lanzhou University, Lanzhou 730000, China}
\affiliation{Lanzhou Center for Theoretical Physics and Key Laboratory of Theoretical Physics of Gansu Province, Lanzhou University, Lanzhou 730000, China.}
\affiliation{Beijing Computational Science Research Center, Beijing 100084, China}

\author{Z. Y. Xie}
\email[]{qingtaoxie@ruc.edu.cn}
\affiliation{Department of Physics,  Renmin University of China,  Beijing 100872,  China}

\begin{abstract}
After decades of debate, now there is a rough consensus that at zero temperature the spin-$1/2$
Heisenberg antiferromagnet on the triangular lattice is three-sublattice $120^\circ$ magnetically ordered,
in contrast to a quantum spin liquid as originally proposed.
However, there remains considerable discrepancy in the magnetization reported among various methods.
To resolve this issue, in this work we revisit this model by the tensor-network state algorithm.
The ground-state energy per bond $E_b$ and magnetization per spin $M_0$ in the thermodynamic limit are obtained with high precision.
The former is estimated to be $E_b = -0.18334(10)$. This value agrees well with that from the series expansion. The three-sublattice magnetic order is firmly
confirmed and the magnetization is determined as $M_0 = 0.161(5)$.
It is about $32\%$ of its classical value and slightly below the lower bound from the series expansion.
In comparison with the best estimated value by Monte Carlo and density-matrix renormalization group, our result is about $20\%$ smaller.
This magnetic order is consistent with further analysis of the three-body correlation.
Our work thus provides new benchmark results for this prototypical model.
\end{abstract}

\maketitle
\section{INTRODUCTION}

One challenging task in modern condensed matter physics is to search for exotic states of matter
both experimentally and theoretically. In this long journey, systems with geometric frustration have emerged as
a flourishing research area. In usual magnets, spins freeze into some periodic patterns upon cooling,
associated with a phase transition from a paramagnetic phase to an ordered phase.
The transition temperature, in comparison with the Curie-Weiss temperature, may be drastically suppressed by geometric frustration.
Actually, in 1973, P. W. Anderson already proposed that some frustrated magnets may remain disordered even at zero temperature,
which is now known as the quantum spin liquid \cite{Anderson1973, Anderson1987, Mila2000, YiZhou2017, Savary2017}.
Ever since then, a large amount of interest has been attracted to search for such exotic states \cite{PALee2008, Balents2010}.
Particularly, in Anderson's original paper \cite{Anderson1973}, the spin-$1/2$ antiferromagnetic
Heisenberg model on the triangular lattice (TAHM) was conjectured to be such a candidate. Moreover, Anderson proposed that its ground state may
be a resonating valence-bond state (RVB) rather than a state with three-sublattice $120^\circ$ magnetic order (TMO) in its classical counterpart.

In the past decades, to clarify the nature of its ground state, TAHM has been extensively studied
by a variety of analytical and numerical methods \cite{Nishi1988, HuseElser1988, Yoshi1991, Bernu1994, Man1998, CapriottiTS1999, Xiang2001, Richter2004,
WeberLMG2006, ZhengFSetal2006, YunokiSorella2006, WhiteChernyshev2007, HeidarianSorellaBecca2009, Chernyshev2009, Kula2013, Suzuki2014, KanekoMoritaImada2014, FarnellGotzeetal2014, LiBC2015, GhioldiMMetal2015, Gotze2016, Iqbal2016, GhioldiGZetal2018}. For example,
Huse and Elser examined this model by variational Monte Carlo \cite{HuseElser1988}. They chose a
trial wavefunction with three-spin terms. By comparing its ground-state energy with that of RVB-type wavefunctions,
they found that the former is energetically favored, and its magnetization is finite, about $68\%$ of its classical value.
On small clusters, exact diagonalization (ED) calculations were performed by several groups but their conclusions
are conflicting \cite{Nishi1988, Bernu1994, Richter2004, Suzuki2014}.
The Green's function Monte Carlo (GFMC) \cite{CapriottiTS1999} and density-matrix renormalization group (DMRG) \cite{WhiteChernyshev2007}
calculations, which were on moderate clusters, concluded the existence of an ordered ground state with a consistent magnetization $M_0\approx{0.205}$.
As far as we know, so far the smallest but finite magnetization reported is $M_0=0.1625(30)$,
obtained by GFMC with fixed node approximation \cite{YunokiSorella2006}.
Now it is mostly believed that the ground state of the TAHM is a TMO state with strongly suppressed magnetization.

However, whereas such progress has been made, the debate has never ceased completely so far.
For example, recent numerical analyses based on bold diagrammatic Monte Carlo \cite{Kula2013} and ED \cite{Suzuki2014} supported the
absence of magnetic order. Moreover, even in those works supporting the existence of TMO,
the discrepancy of the magnetization is quite large, with its value
ranging from 0.1625(30) to 0.36 \cite{YunokiSorella2006,  WeberLMG2006}. And finally,
from the experimental perspective, various compounds with triangular geometry have been synthesized and fingerprints of quantum spin
liquids were reported \cite{Zhou2011, Li2013}, but their nature remains controversial.
As a prototypical model with geometric frustration, precise understanding of the TAHM is important and necessary.
In particular, an accurate estimate of the magnetization may help us
to understand related experiments and serve as a benchmark for newly developed numerical algorithms.
It is fair to say that the present knowledge remains unsatisfactory and thus calls for further studies on this model.

For this purpose, we revisit this model by tensor-network state (TNS) method \cite{Nig1996, Nishino2001, PEPS2004}
which is under rapid development and has drawn great attention due to its successful applications in strongly-correlated
condensed matter physics \cite{TO2013, tJ2014, Kagome2017}, statistical physics \cite{HOTRG2012, CW2014, MBL2015},
quantum field theory \cite{CMPS2010, LQCD2013, CTNS2019}, and machine learning \cite{ML2018, ML2020}, etc.
To be specific, the TAHM is described by the Hamiltonian
\begin{eqnarray}
	\mathcal{H} & = & J\sum_{\langle ij \rangle} {\hat{\bf{S}}}_{i}\cdot{\hat{\bf{S}}}_{j}, \label{HAM}
\end{eqnarray}
where $J>0$ is the antiferromagnetic coupling. Hereafter we set $J=1$ as the energy unit. ${\hat{\bf{S}}}_i$ is the spin operator at site $i$.
$\langle \cdots \rangle$ means a summation over the nearest-neighbor pairs. We use the the projected entangled simplex state (PESS)
ansatz \cite{PESS2014} to represent the ground-state wavefunction, and employ the corner transfer-matrix renormalization group (CTMRG)
method \cite{CTMRG1996, CTMRG2009, tJ2014} to estimate the physical quantities, such as $E_b$, $M_0$,
and many-body correlation \cite{BeiBook2019}.

The rest of the paper is organized as follows. In Sec. II, we introduce some details of the algorithm employed in our work.
The numerical results for $E_b$, $M_0$ and many-body correlation are present in Sec. III. In Sec. IV, we summarize our work.

\section{METHODS}
Frustration in TAHM makes it difficult to be investigated with traditional numerical methods such as Monte Carlo, which suffers from the infamous sign problem and strong finite-size effect. Generally, the TNS method is free of the sign problem and can study this model in the thermodynamic limit directly by assuming a translationally invariant wavefunction. Therefore, it is drawing increasing attention nowadays.

\begin{figure}
\includegraphics[width=0.45\textwidth]{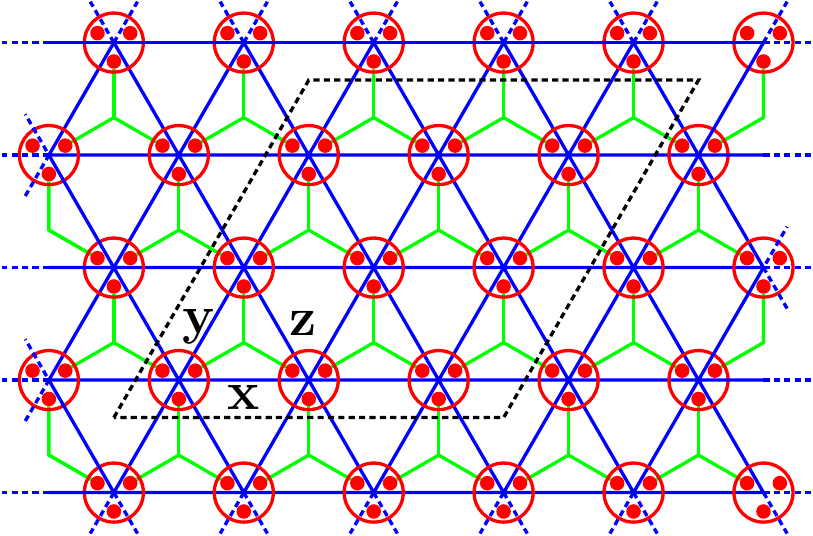}
\caption{Schematic diagram of the PESS wavefunction ansatz on the infinite triangular lattice.
	The blue lines are the bonds of the lattice, which are marked by $x, y$ and $z$, respectively.
	The green lines represent virtual bonds of the wavefunction.
	The tensors sitting at the center of the triangles are the simplex tensors $S$, and the tensors covered by red circle are the projection tensors $A$.
	The rhombus with dashed lines marks a $3\times 3$ unit cell of the trial wave function.
	The physical indices are perpendicular to the plane and not shown here.}
	\label{3-pess}
\end{figure}

In the TNS family, PESS is a wavefunction ansatz \cite{PESS2014} generalized from the popular
projected entangled pair state (PEPS) ansatz \cite{PEPS2004}, and is believed to be suitable for frustrated systems.
In this work, the PESS ansatz is defined as
\begin{equation}
|\Psi\rangle = \sum_{\{\sigma\}}\mathrm{Tr}(...S^{(\mu\nu)}_{i_{\mu\nu}j_{\mu\nu}k_{\mu\nu}}
    A^{(\lambda\omega)}_{i_{\lambda\omega}j_{\lambda\omega}k_{\lambda\omega}}
    [\sigma_{\lambda\omega}]...)|...\sigma_{\lambda\omega}...\rangle
\label{Eq:3-pess}
\end{equation}
which is illustrated in Fig.~\ref{3-pess}. Here $(\mu,\nu)$ denotes the location of the upward triangles,
and $(\lambda,\omega)$ denotes the location of the lattice sites. A rank-3 simplex tensor $S$ is defined
at the center of each upward triangle, and a rank-4 projection tensor $A$ is defined at each lattice site.
$\{i, j, k\}$ and $\{\sigma\}$ are the virtual indices and physical basis associated with the tensors, respectively.
The two virtual indices associated with the same bond take the same values. $\mathrm{Tr}$ is over all the repeated
virtual indices and $\sum$ is over all the basis configurations.

To employ the translational invariance, we use a $3\times 3$ periodicity, which means that
\begin{equation}
S^{(\mu,\nu)} = S^{(\mu+3m, \nu+3n)}, \quad A^{(\lambda,\omega)} = A^{(\lambda+3m, \omega+3n)}
\end{equation}
where $m,n$ are integers.
In other word, we totally have 9 different $S$ and 9 different $A$ in the ansatz~(\ref{Eq:3-pess}).
The corresponding unit cell is illustrated by a dashed rhombus in Fig.~\ref{3-pess}.

It is known that the bond dimension, $D$, which is the maximal value of the virtual indices, controls the number of independent parameters
and thus the numerical accuracy. In this work, $D$ is up to 13. The ground-state wavefunction
is optimized by simple update algorithm \cite{SU1D2007, SU2D2008}.
Though the full update strategy \cite{FU2014} might be more accurate, it is much more costly.
To verify the result, we compared the magnetizations at $D = 6$ so that the full update and the recent automatic 
differentiation \cite{AD} can be performed. The simple update approach gives an estimation about 0.2448. Starting from such a wave function, 
full update and automatic differentiation \cite{CMPD6} can further reduce the magnetization down to 0.2395 and 0.2382, respectively. 
The difference among these results is of the order $10^{-3}$.
In viewing of the computational cost, we choose the more efficient simple update scheme in this work. The numerical accuracy can be remedied by larger $D$.
In order to avoid the bias and reduce the Trotter error, we started from a wavefunction randomly generated in complex field,
and gradually reduced the Trotter step $\tau$ from a large value, say $0.2$. The final $\tau$ is smaller than $10^{-3}$,
which turns out to be sufficiently small to estimate the magnetization of TAHM.

\begin{figure}[htbp]
\centering
\subfigure[]{
\begin{minipage}[t]{0.48\linewidth}
\centering
\includegraphics[width=4.6cm]{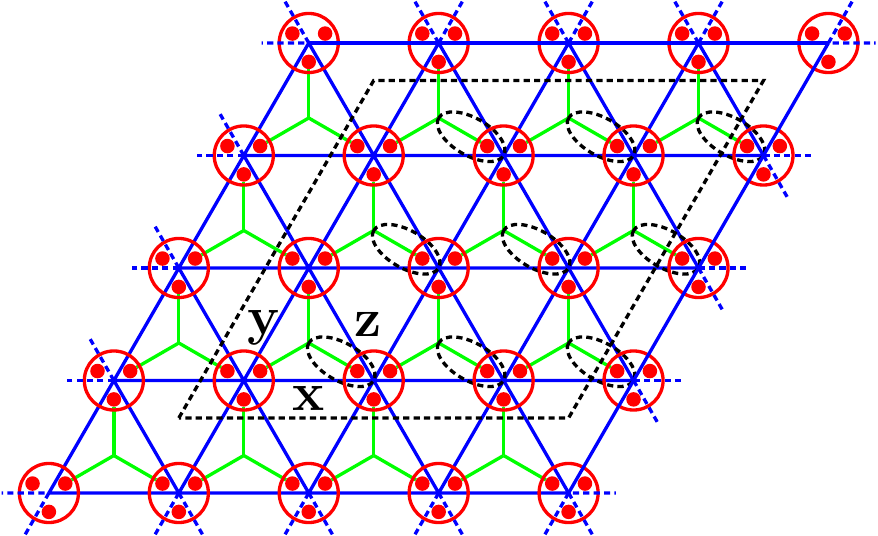}
\end{minipage}
\label{Fig:CTM1}}
\subfigure[]{
\begin{minipage}[t]{0.45\linewidth}
\centering
\includegraphics[width=3.4cm]{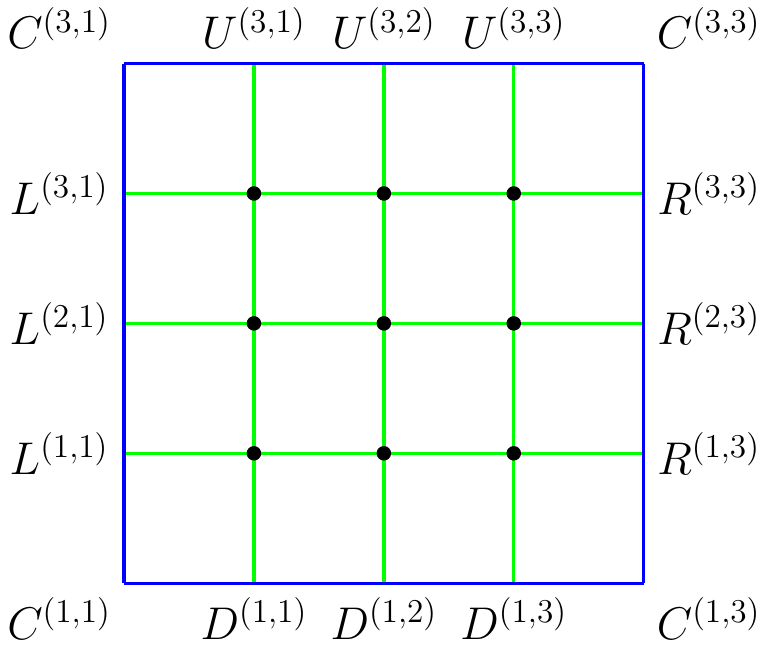}
\end{minipage}
\label{Fig:CTM2}}
\caption{(a) Converting the tensor network skeleton from the honeycomb lattice to a square lattice by one-step contraction, e.g., in the direction of the bonds surrounded by dashed ellipses. (b) The $3\times 3$ unit cell obtained after deformation in the reduced
	network $\langle\Psi|\Psi\rangle$. Here the environment tensors of the unit cell are shown explicitly, e.g., $L^{(3,1)}$ are the edge tensor
	associated with the left of $T^{(3,1)}$.}
\label{Fig:CTM}
\end{figure}

Physical observables are calculated via the CTMRG method, which was developed for an arbitrary unit cell on the square lattice \cite{tJ2014}.
In Fig.~\ref{3-pess}, we show the PESS ansatz defined on honeycomb skeleton. Firstly, we formally deform the skeleton to a square by
simply combining $S$ with $A$ together to form a single tensor $T$, e.g.,
\begin{equation}
T^{(\mu\nu)}_{k_1k_2i_1i_2}[\sigma] = \sum_{j}S^{(\mu\nu)}_{i_1jk_1}A^{(\mu,\nu+1)}_{i_2jk_2}[\sigma]
\end{equation}
This is done in all the upward triangles coherently, as illustrated in Fig.~\ref{Fig:CTM1}.
Hence, the reduced network $\langle\Psi|\Psi\rangle$, which appears in expectation value calculation, see Eq.~(\ref{EBOND}) and (\ref{MSITE}),
can be represented as a two-dimensional tensor network with a $3\times 3$ periodicity, as illustrated in Fig.~\ref{Fig:CTM2},
and then the standard CTMRG method can be applied directly to contract the network. Finally the local physical observables can be
calculated efficiently from the local environment tensors $\{L, R, U, D, C\}$. Similarly, the bond dimension $\chi$ of the
environment tensors is a tunable parameter which controls the accuracy in CTMRG. In our calculation, the maximal $\chi$ is no less than $D^2$
to ensure a reliable result \cite{NTS2017}.

\section{RESULTS}
\subsection{Ground-State Energy}
The ground-state energy usually serves as a key criterion for trial wavefunctions,
particularly in the variational Monte Carlo simulations. This is exactly how Huse and Elser excluded the
quantum spin liquid ground state in TAHM \cite{HuseElser1988}. From this aspect, an accurate estimate of the ground-state energy is important.
Therefore, firstly we need to check whether our numerical results are reliable, by comparing the ground-state energy
with that in previous works.

The ground-state energy for a given bond $\langle ij \rangle$ is given by
\begin{eqnarray}
	E_{\langle ij \rangle}=\frac{\langle \Psi|{\hat{\bf{S}}_i}\cdot{\hat{\bf{S}}_j}|\Psi\rangle}{\langle \Psi|\Psi \rangle} \label{EBOND}
\end{eqnarray}
where $|\Psi\rangle$ is the PESS representation of the ground-state wavefunction, see Eq.~(\ref{Eq:3-pess}).
Since our system is translationally invariant, the bond energy $E_b$ can be estimated by averaging $E_{\langle ij \rangle}$ over all bonds in one unit cell.

As stated in the previous section, the accuracy of the wavefunction is controlled by $D$,
and that of the expectation is controlled by $\chi$. Therefore, to obtain accurate results for a given $D$,
the expectation values are calculated with a series of $\chi$ in which the largest one is no less than $D^2$,
and then extrapolated as $\chi\rightarrow\infty$.

\begin{figure}[!ht]
	\includegraphics[width=0.95\columnwidth, clip]{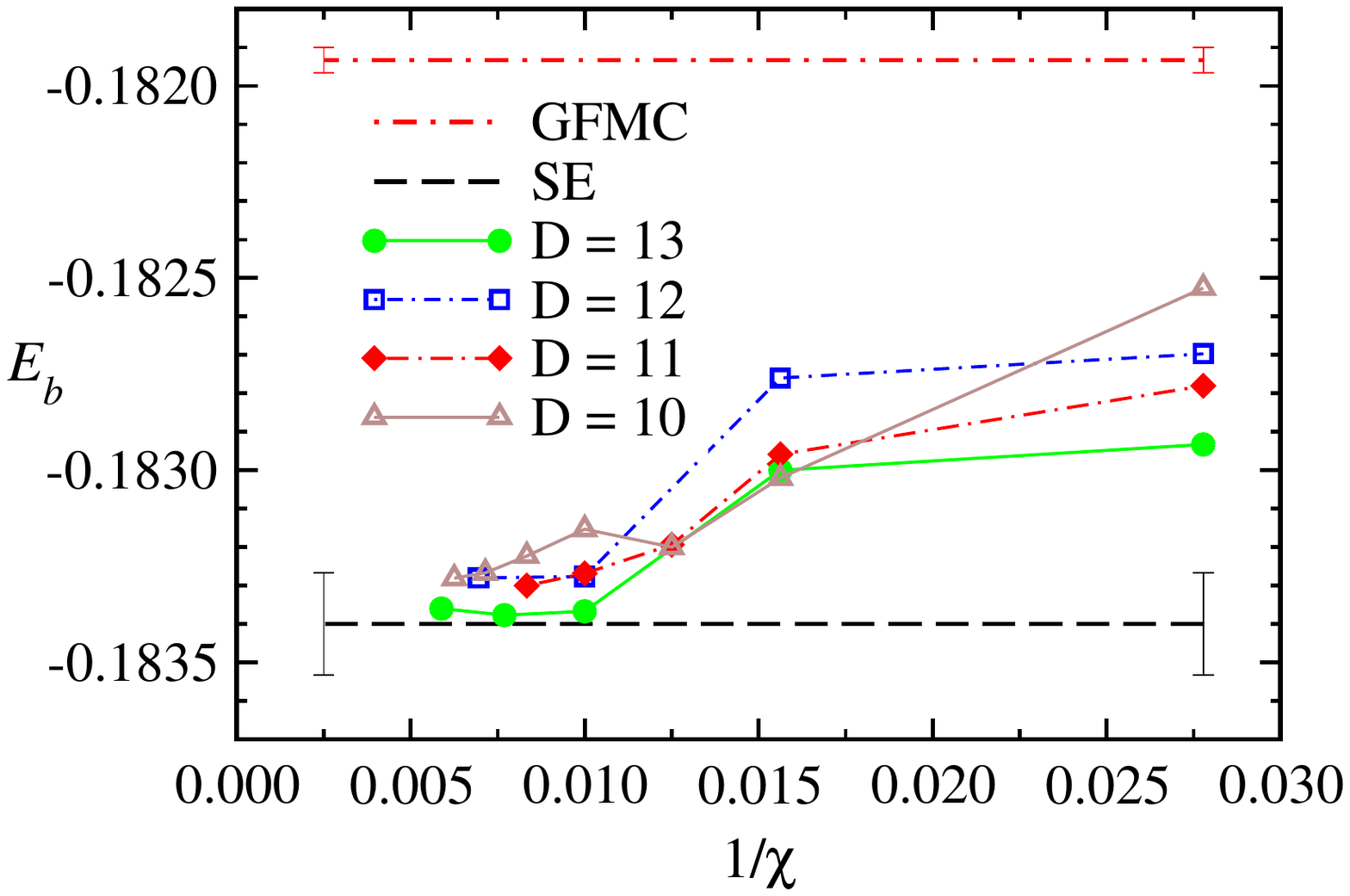}\\
	\caption{The ground-state energy $E_b$ for $D=10,11,12$ and $13$ is plotted as a function of $\chi$. The numerical error is on the
	fifth digit and they are not shown for a clear vision. The data obtained by GFMC \cite{CapriottiTS1999}
	and SE \cite{ZhengFSetal2006} are also shown for comparison. Our data are obviously
	below that from GFMC but agrees well with that by SE (within the error bar).}
	\label{FIG_Energy}
\end{figure}

For the smallest $D = 4$ in our simulations, the ground-state energy is $E_b = -0.18226(9)$, which is already
lower than that obtained by GFMC \cite{CapriottiTS1999}, -0.18193(3), and is also lower than that obtained by infinite-PEPS calculation whose ansatz is defined on the decorated square latice \cite{DSL} with $D = 4$, -0.1813.   
To provide an intuitive impression, in Fig.~\ref{FIG_Energy}, we plot $E_b$ as a function
of $1/\chi$ for $D = 10, 11, 12$ and $13$. It seems that $E_b$ depends very weakly on $D$ when $\chi$ becomes large,
and all the data points are well below those from GFMC.

As $\chi$ increases, $E_b$ roughly decreases monotonically, but they oscillate in a small interval as a function of $D$.
For the data points with largest $\chi$ in Fig.~\ref{FIG_Energy}, $E_b$ is between $-0.18328$ and $-0.18336$.
With the available $\chi$, this non-monotonic behavior with regard to $D$ makes it difficult to extrapolate our data
and hinder us to obtain more accurate results. As a compromise, we firstly extrapolate the data for $D = 10,11,12$u
and $13$ to the infinite $\chi$ limit, respectively, and then average them.
Our final result is $E_b = -0.18334(10)$, which agrees well with that obtained by
the series expansion (SE) \cite{ZhengFSetal2006} and the coupled cluster method \cite{Gotze2016}.

\begin{table}
	\begin{tabular}{c|cc|cc|cc}
		\hline
		\hline
		Method & $E_b$ & & & $M_0$ && Year \\
		\hline
		 this work      &    -0.18334(10)    &&&     0.161(5)   && 2020\\
		 SB+1/N \cite{GhioldiGZetal2018}     &        ---         &&&     0.224      && 2018\\
		 DMRG \cite{Iqbal2016}      &   -0.1837 (7)              &&&      ---        && 2016\\
		 CC \cite{Gotze2016}     &    -0.1838         &&&     0.21535   && 2016 \\
		 SB \cite{GhioldiMMetal2015}         &        ---         &&&     0.2739     && 2015\\
		 SWT \cite{GhioldiMMetal2015}        &        ---         &&&     0.2386     && 2015\\
		 SE  \cite{GhioldiMMetal2015}         &        ---         &&&     0.198(34) && 2015 \\
		 CC \cite{LiBC2015}       &     -0.18403(7)    &&&     0.198(5)   && 2015 \\
		 CC \cite{FarnellGotzeetal2014}      &     -0.1843        &&&     0.1865     && 2014 \\
		 VMC \cite{KanekoMoritaImada2014}      &     -0.18163(7)    &&&     0.2715(30) && 2014 \\
		 SWT \cite{Chernyshev2009}   &     -0.18228       &&&     0.24974   && 2009 \\
		 VMC \cite{HeidarianSorellaBecca2009}    &     -0.18233(3)    &&&     0.265     && 2009 \\
		 DMRG \cite{WhiteChernyshev2007}          &        ---         &&&     0.205(15)  && 2007\\
		 FN \cite{YunokiSorella2006}            &     -0.17996(1)    &&&     0.1625(30) && 2006\\
		 FNE \cite{YunokiSorella2006}           &     -0.18062(2)    &&&     0.1765(35) && 2006 \\
		 SE   \cite{ZhengFSetal2006}           &     -0.18340(13)   &&&     0.19(2)    && 2006\\
		 VMC \cite{WeberLMG2006}          &     -0.1773(3)     &&&     0.36       && 2006 \\
         ED \cite{Richter2004}          & -0.1842            &&& 0.193      && 2004 \\
		 DMRG \cite{Xiang2001}          &     -0.1814        &&&      ---          && 2001 \\
		 GFMC \cite{CapriottiTS1999}    &       -0.18193(3)  &&&    0.205(10)      && 1999\\
		\hline
		\hline
	\end{tabular}
\caption{$E_b$ and $M_0$ obtained by various methods are shown for comparison. SB, CC, SWT, VMC, FN, and FNE denote Schwinger boson mean field theory, coupled cluster, spin-wave theory, variational Monte Carlo, fixed node, and fixed node with effective Hamiltonian, respectively.}
\label{tab:mag}
\end{table}

In Tab.~\ref{tab:mag}, we summarize some recent works for comparison. These data indicate that our PESS wavefunction represents a
good approximation of the ground state of TAHM.

\subsection{Magnetization}
The main debate about this model is whether the ground state is a TMO state or a quantum spin liquid.
From Tab.~\ref{tab:mag}, we can see that, even in those works advocating TMO, the magnetization $M_0$ differs significantly.
For example, if the error bar is taken into account, the low bound given by SE \cite{GhioldiMMetal2015} is
smaller than half of that given in Ref. \cite{WeberLMG2006}.
This motivates us to calculate the magnetization in this work.
\begin{figure}[!ht]
	\includegraphics[width=0.95\columnwidth, clip]{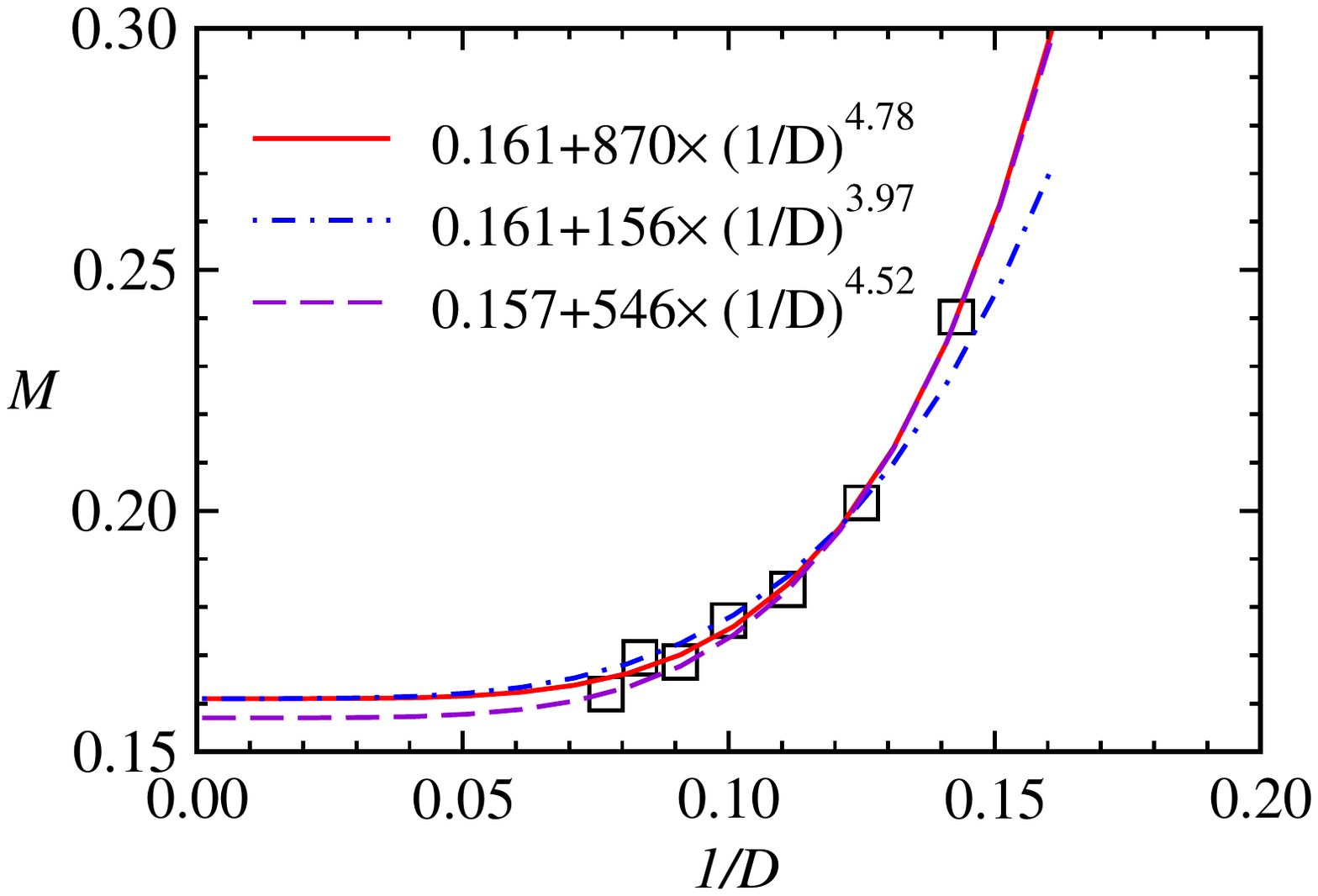}
	\caption{$M$, marked as ${\boldsymbol{\square}}$, is plotted as a function of $1/D$.
	         All the data points have already been extrapolated to the infinite-$\chi$ limit. The lines are different numerical fittings:
		 solid line is obtained from all $D$, while dot-dashed line and dashed line are obtained from even and odd $D$ only, respectively.}
	\label{MAG}
\end{figure}
Given the ground state $|\Psi\rangle$, three components of the magnetization vector ${\vec{M}_i}$ at the site $i$ are given by
\begin{eqnarray}
        M_{i}^{\alpha}  =  \frac{\langle \Psi| \hat{\bf{S}}_i^\alpha | \Psi \rangle}{\langle \Psi|\Psi\rangle} \label{MSITE}, \quad\quad \alpha = x, y, z
\end{eqnarray}
from which the magnetization at site $i$ reads
\begin{eqnarray*}
M_i=\sqrt{(M_i^x)^2+(M_i^y)^2+(M_i^z)^2},
\end{eqnarray*}
and the relative angles between neighbouring spins are immediately available.
In the calculation, we found that the magnetization is almost independent of the sites. For simplicity, hereafter we show only the overall magnetization $M$,
which is obtained by averaging over all the $M_i$ within one unit cell. Similar to the calculation of $E_b$, for a given $D$, we extrapolate $M$ as a function of $1/\chi$ to the infinite $\chi$ limit.

The results for $D$ from $7$ to $13$ are illustrated in Fig.~\ref{MAG}. We notice that for $D = 9$,
our result is already smaller than most of recent results, see Tab.~\ref{tab:mag}.
Clearly, it shows that $M$ decreases roughly as a monotonic function of $1/D$. To get a more accurate estimate,
we try to fit them with two typical formulae. One is an power-law formula, i.e., $M = M_0 + a\times(1/D)^{b}$, yielding $M_0 = 0.161$.
The other is an exponential formula, i.e., $M = M_0 + a\times\exp(-bD)$, with $M_0 = 0.164$ for the best fit.

With a careful inspection of Fig.~\ref{MAG}, we notice that there is a tiny even-odd oscillation in the magnetization as a function of $D$,
which suggests us fit the magnetization for even and odd $D$ separately. Using the power-law formula,
we obtain $M_0 = 0.161$ and $M_0 = 0.157$ for even and odd $D$, respectively. Defining the error bar as the
standard deviation among the four different $M_0$ obtained above, we conclude that $M_0 = 0.161(5)$,
which is very close to the lower bound obtained by SE~\cite{GhioldiMMetal2015, ZhengFSetal2006}.
One may notice that this value is also very close to that in Ref.~\cite{YunokiSorella2006}, but their ground-state energy is obviously not optimal.
More details can be found in Tab.~\ref{tab:mag}.

We would like to emphasize that the magnetization we obtained is slightly smaller than $1/3$ of its classical value.
In particular, it is smaller than all that obtained in previous works. On one hand, such a small magnetization requires
a careful finite-size analysis to obtain a quantitatively reliable estimation in numerical calculations such as
the ED, DMRG, and Monte Carlo. On the other hand, generally, TNS method usually tends to overestimate the magnetization
in frustrated systems when $D$ is finite~\cite{Kagome2017}. This suggests that probably our smallest result for finite $D$ is the upper bound of the
magnetization. Therefore, it is quite likely that $M_0$ has been overestimated in previous works.

\begin{table}
	\begin{tabular}{cccccccc}
		\hline
		\hline
		\multirow{2}{*}{($\mu, \nu$)}&\multicolumn{3}{c}{$D=4, \chi=32$}&\multicolumn{4}{c}{$D=13, \chi=170$} \\ \cline{2-4} \cline{6-8}
		&x&y&z& &x&y&z\\
		\hline
		(1, 1) &120.004 &120.000 &119.996 &  &120.010 &119.988 &120.002\\
		(1, 2) &119.999 &120.000 &120.001 &  &119.985 &119.994 &120.021\\
		(1, 3) &119.997 &120.000 &120.003 &  &120.005 &119.993 &120.002\\
		(2, 1) &120.004 &120.000 &119.996 &  &120.005 &120.012 &119.983\\
		(2, 2) &119.999 &120.000 &120.001 &  &119.992 &120.013 &119.994\\
		(2, 3) &119.997 &120.000 &120.003 &  &120.004 &120.013 &119.983\\
		(3, 1) &120.004 &120.000 &119.006 &  &120.003 &120.001 &119.996\\
		(3, 2) &119.999 &120.000 &120.001 &  &119.993 &120.993 &119.986\\
		(3, 3) &119.997 &120.000 &120.003 &  &120.004 &119.994 &120.998\\
		\hline
		\hline
	\end{tabular}
\caption{Angles (in unit of degree) of the magnetization vectors between nearest neighbors are shown.
	Location of the upward triangles in the unit cell is listed explicitly.
$x,y$ and $z$ are the three directions in the triangle lattice, as marked in Fig. \ref{3-pess}. They indicate the corresponding bonds of the triangle here.}
\label{tab:angle}
\end{table}
In Tab.~\ref{tab:angle}, we present the data of the angles between all the nearest neighbors in the unit cell,
for two sets of parameters, i.e., $D = 4$ with $\chi = 32$ and $D = 13$ with $\chi = 170$.
It shows that: (I) the $120^\circ$ angles between nearest neighbors are almost perfect, in the sense that the largest
error bar is as small as $0.021^\circ$ with $D$ up to 13, (II) in contrast to the magnetization, the angles are almost
independent of $D$ and $\chi$, as long as they are not too small. Therefore, we can safely conclude the existence of the TMO.

\subsection{Larger Unit Cell}
\begin{figure}[ht]
	\includegraphics[width=0.45\textwidth, clip]{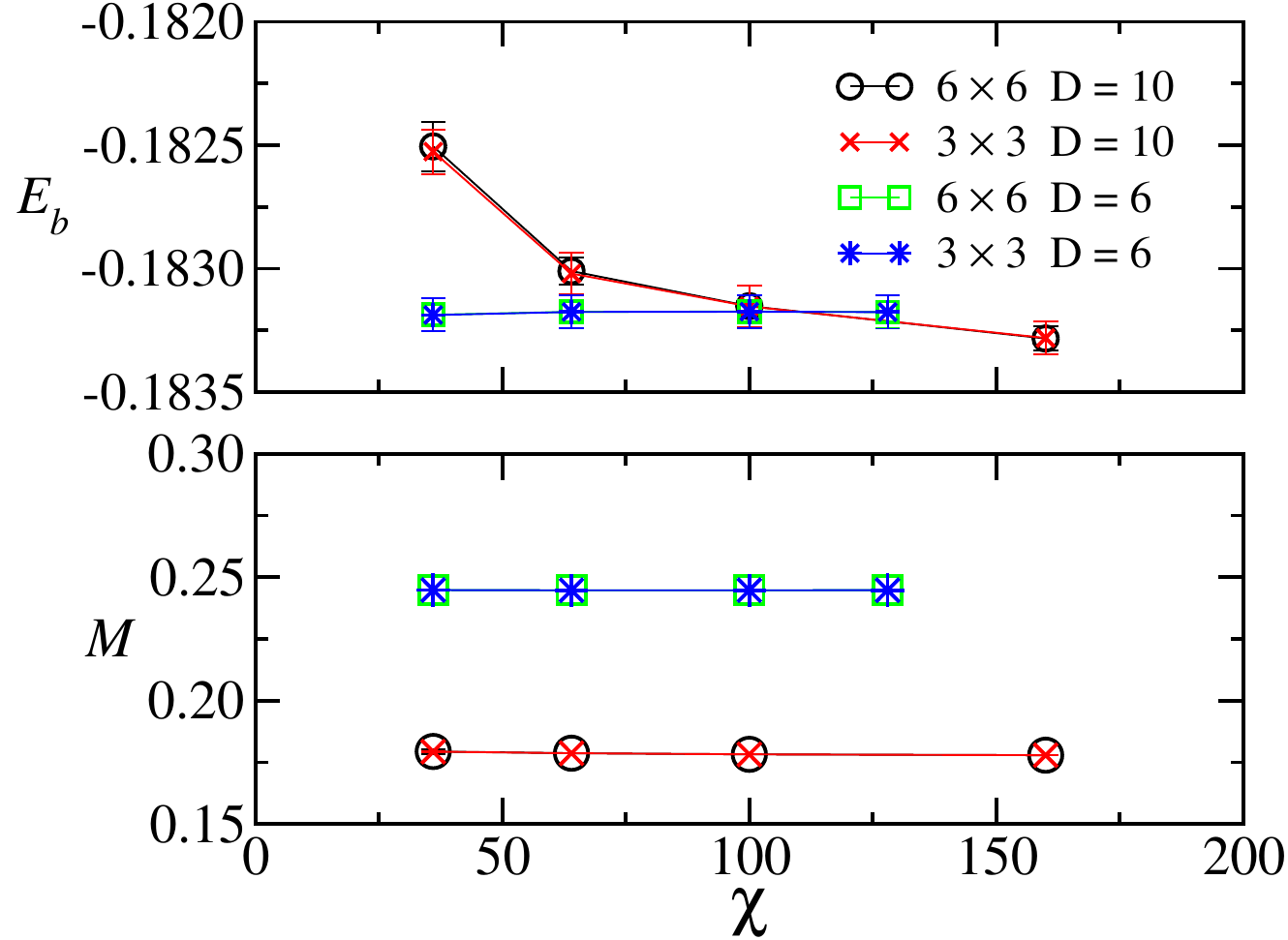}
	\caption{$E_b$ and $M$ are shown as a function of $\chi$ for $6\times6$ and $3\times 3$ unit cells. Our results show
	that they agree well, suggesting that the $3\times 3$ unit cell is large enough for TAHM.}
	\label{Fcluster}
\end{figure}

The result of TNS simulation might also depend on the size of the unit cell, thus we need to check
whether the unit cell we used in the wavefunction ansatz is sufficiently large. For this purpose,
we compare our results from the $3\times 3$ unit cell with those from the $6\times 6$ unit cell.
In Fig. \ref{Fcluster}, we plot $E_b$ and $M$ as a function of $\chi$ for
$D = 10$ and $6$. The data is in excellent agreement for the two different unit cells,
and the differences at all data points are negligible compared to the error bar.
This suggests that the $3\times 3$ unit cell in our work is already large enough for TAHM.

\subsection{Many-Body Correlation}
 The motivation to study the many-body correlation in this model comes from two perspectives.
 On one hand, the existence of TMO indicates that in each triangle there is probably some three-body
 correlation that is essentially different from the two-body correlation. Actually, this is one reason why we use PESS ansatz to study this model.
 On the other hand, from the view of quantum information, for mixed many-body states, generally the total correlation leaks more information than the part peculiar to quantum states only, i.e., entanglement, which has no classical counterpart \cite{BeiBook2019}. What's more, though PESS is believed to be able to capture the many-body correlation better, there has no direct numerical evidence yet to demonstrate the existence of such correlation in the obtained wavefunction.
 Therefore, the frustrated TAHM offers such an opportunity to study the many-body correlation, especially the three-body correlation in a triangle.

To be specific, we envisage that the three spins $\{\sigma_a,\sigma_b,\sigma_c\}$ in a triangle comprise a mixed quantum state,
which can be characterized by the reduced-density matrix $\rho^{(3)}$ defined below
\begin{equation}
\rho^{(3)}_{II'} = \sum_{J}|\Psi_{IJ}\rangle\langle\Psi_{I'J}|
\end{equation}
where $I$ and $J$ denote the composite physical indices corresponding to $\{\sigma_a,\sigma_b,\sigma_c\}$ and
the rest spins in the ground state, respectively. Similarly we can define $\rho^{(1)}$ for one spin and $\rho^{(2)}$
for a pair of spins sharing one bond.

Once the three kinds of mixed states are defined, we can calculate the von Neumann entropies, $S = - \mathrm{Tr}\rho\ln\rho$, for these states. For simplicity, we use $S_i$, $S_{ij}$, $S_{ijk}$ to denote the entropies corresponding to spin $\sigma_i$, spin pair $\{\sigma_{i},\sigma_{j}\}$ and spin simplex $\{\sigma_i,\sigma_j,\sigma_k\}$, respectively, with $i,j,k = a, b, c$. Then we measure the correlations in this small triangle through the following quantities defined below
\begin{eqnarray}
I_{a} &=& S_a  \nonumber \\
I_{ab} &=& S_a + S_b - S_{ab} \nonumber \\
I^{(3)} &=& S_a + S_b + S_c - S_{abc}
\label{eq:VE}
\end{eqnarray}
where $I_{ab}$ and $I^{(3)}$ are the two-body and three-body mutual information which are used to measure the total correlation for a general quantum system \cite{BeiBook2019}, respectively. Other terms can be obtained similarly. Moreover, the true tripartite correlation $I^{(3)}_{tr}$, which is more relevant in this context, can be identified from $I^{(3)}$ by excluding the pair correlation contributions, i.e.,
\begin{equation}
I^{(3)}_{tr} = I^{(3)} - I_{ab} - I_{bc} - I_{ca}
\label{eq:VEt}
\end{equation}

\begin{figure}[ht]
	\includegraphics[width=0.45\textwidth, clip]{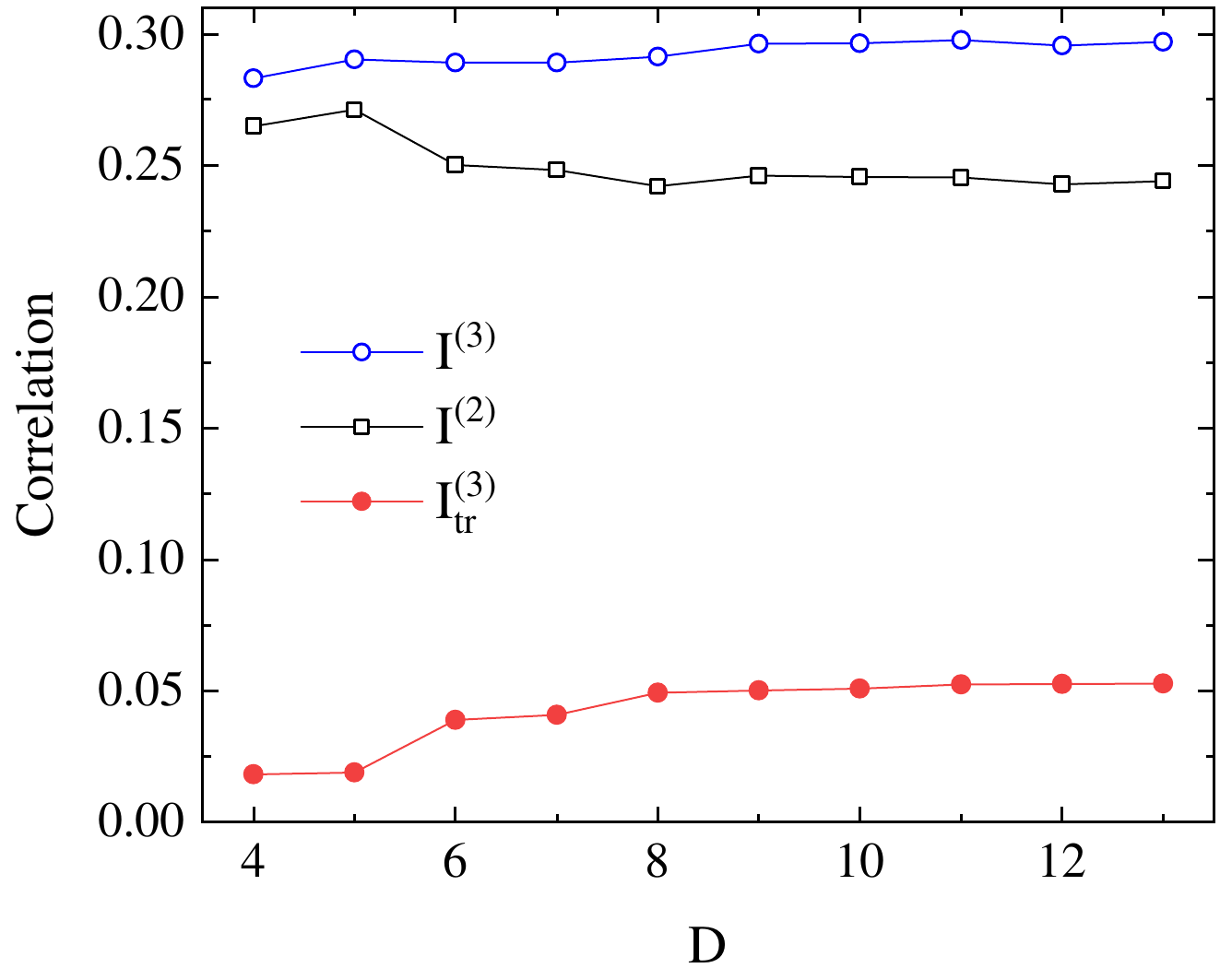}
	\caption{Correlation measured by mutual information in one triangle of the ground-state wavefunction. Here, $I^{(2)}$ denotes the total pair correlation, namely $I^{(2)} = I_{ab} + I_{bc} + I_{ca}$. See Eq.~(\ref{eq:VE}) and (\ref{eq:VEt}).}
	\label{Fig:Correlation}
\end{figure}

The obtained results are shown in Fig.~\ref{Fig:Correlation}. We can see clearly that in this frustrated system, as $D$ becomes larger, pair correlation becomes weaker, while simplex correlation becomes stronger. More importantly, it shows that as $D$ increases, the true tripartite correlation $I^{(3)}_{tr}$ becomes more and more significant, which coincides with the fact that the TMO can be argued to have imposed a global constrain on the three spins simultaneously, not just a local constrain on each pair in the triangle. This makes us more confident that the ground state should be of TMO, and that the PESS wavefunction can indeed grasp well the many-body correlation in this model.

\section{Summary}
In summary, using tensor-network algorithms with PESS-type trial wave function, we have studied the spin-$1/2$ antiferromagnetic
Heisenberg model on the triangular lattice. This wavefunction was optimized by the simple update imaginary-time evolution method,
and the expectation values were estimated by the multi-sublattice CTMRG algorithm. By comparing the ground-state energy to
that in other works, we confirmed that the wavefunction converges to the ground state and it is a TMO state.
In particular, the magnetization is $M_0=0.161(5)$, which is smaller than that reported in previous calculations like GFMC, DMRG.
Although frustration and quantum fluctuation do introduce some unusual properties into the model, such as roton-like excitations \cite{ZhengFSetal2006},
its ground state remains magnetically ordered.
This result is consistent with the correlation analysis, which shows that as $D$ increases, the two-body correlation
becomes weaker gradually, while the three-body correlation becomes increasingly significant. In viewing of the experience
that TNS method, especially when simple update strategy is used, may tend to overestimate the magnetization of frustrated systems
a little bit for a finite $D$, (as evidenced by the comparison for $D=6$ in the main text, for example), we believe that our work provides new benchmark results for this model. \\

\section{Acknowledgement}

J. Z. is supported by the National Natural Science Foundation of China (Grant No. 11874188),
H. L. is supported by the National Natural Science Foundation of China (Grant No. 11674139, 11834005),
Z. Y. Xie is supported by the National R$\&$D Program of China (Grants No. 2017YFA0302900 and No. 2016YFA0300503),
the National Natural Science Foundation of China (Grants No. 11774420), and the Research Funds of Renmin University
of China (Grants No. 20XNLG19). We thank Hai-Jun Liao and Hai-Yuan Zou for helpful discussions about automatic differentiation and PEPS calculations. Qian Li and Hong Li contributed equally to this work.


\begin{thebibliography}{99}
\bibitem{CMPD6} The full update and automatic differentiation were performed with environment bond dimension $\chi = 50$, and the result was roughly converged for $D = 6$.
\bibitem{DSL} Though the decorated square lattice ansatz is not a good ansatz for triangular lattice due to the likely 3-fold rotational symmetry breaking, we have not tried the PEPS ansatz defined on the original triangular lattice because of the extremely high cost, since coordination number there is 6 which is too large for tensor-network calculations.
\end{thebibliography}
\end{document}